\documentclass[11pt,preprint]{aastex}
\usepackage{psfig}

\newcommand\simlt{\lower.5ex\hbox{$\; \buildrel < \over \sim \;$}}
\newcommand\simgt{\lower.5ex\hbox{$\; \buildrel > \over \sim \;$}}

\slugcomment{accepted for publication in the Astrophys. J. Letters (June 2000)}

\shorttitle{R. Baptista \& M. S. Catal\'an}
\shortauthors{A spiral structure in EX Dra}

\begin{document}

\title{A spiral structure in the disk of EX Draconis on the rise to 
		outburst maximum}
\author{Raymundo Baptista}
\affil{Departamento de F\'\i sica, Universidade Federal de Santa Catarina,
		Campus Trindade, 88040-900, Florian\'opolis - SC, Brazil}
\email{bap@fsc.ufsc.br}
\and
\author{M. S. Catal\'an}
\affil{Department of Physics, Keele University, Keele, Staffordshire,
       ST5 5BG, UK}
\email{msc@astro.keele.ac.uk}

\begin{abstract}

We report on the R-band eclipse mapping analysis of high-speed photometry
of the dwarf nova EX Dra on the rise to the maximum of the November 1995
outburst.
The eclipse map shows a one-armed spiral structure of $\sim 180\degr$ 
in azimuth, extending in radius from $R\simeq 0.2$ to $0.43\;R_{\rm L1}$
(where $R_{L1}$ is the distance from the disk center to the inner 
Lagrangian point), that contributes about 22 per cent of the total flux
of the eclipse map.
The spiral structure is stationary in a reference frame co-rotating
with the binary and is stable for a timescale of at least 5 binary
orbits.
The comparison of the eclipse maps on the rise and in quiescence 
suggests that the outbursts of EX Dra may be driven by episodes of 
enhanced mass-transfer from the secondary star.
Possible explanations for the nature of the spiral structure are discussed.

\end{abstract}


\section{Introduction}

Dwarf novae are close binaries in which a late type star 
(the secondary) overfills its Roche lobe and transfers matter 
to a companion white dwarf via an accretion disk. 
They show recurrent outbursts of 2--5 magnitudes on timescales of days
to several months, resulting either from an instability in the mass 
transfer from the secondary (the MTI model), or from a thermal 
instability in the accretion disk (the DI model) which switches the 
disk from a low to a high-viscosity regime (Warner 1995 and references 
therein). 
Observations of dwarf novae at the early phases of the outburst offer 
good prospects to discriminate between the two competing models. 
The MTI model predicts that the accretion disk should shrink at the 
beginning of the outburst due to the sudden accretion of material with
low angular momentum while the luminosity of the bright spot should 
increase in response to the higher mass-transfer rate (e.g. Warner 
1995). No such effects are expected in the DI model. 
Most of the few current observations seem to support the DI model 
(e.g., Vogt 1983; Rutten et~al. 1992a), although recent results 
suggest an MTI origin for the outbursts in the intermediate polar 
EX Hya (Hellier et~al. 2000).

Spiral shocks have long been proposed as a possible mechanism for 
transport of angular momentum in accretion disks (e.g., Sawada, Matsuda
\& Hachisu 1986; Savonije, Papaloizou \& Lin 1994) and may be the key,
together with magnetic viscosity (Hawley, Balbus \& Winters, 1999), 
in understanding the anomalous viscosity mechanism responsible for 
the spiraling inwards of the disk material.
The recent discovery of spiral shocks in the accretion disk of the
dwarf nova IP~Pegasi in outburst (e.g., Steeghs, Harlaftis \& Horne
1997; Baptista, Harlaftis \& Steeghs 2000a) confirmed the results of 
hydrodynamical simulations (e.g., Stehle 1999) and boosted the 
research in this topic.
Two-armed spiral shocks are excited in the outer regions of the disk 
by the tides raised by the secondary star when the outbursting 
accretion disk extends far enough into the Roche Lobe.

EX Draconis is a long period ($P_{orb}=5$ hr) eclipsing dwarf nova,
the outbursts of which have typical amplitudes of 2.0 mag, duration
of $\simeq 10$ days, and recurrence timescale of about 20 days 
(Baptista, Catal\'an \& Costa 2000b; hereafter BCC).
The analysis of a set of light curves along the outburst cycle led BCC
to conclude that the outbursts do not start in the outer disk regions.
(i.e., they are inside-out outbursts).
Recently, Joergens, Spruit \& Rutten (2000) found evidence of spiral
shocks in the accretion disk of EX Dra from Doppler tomography close 
to outburst maximum.

In this letter we report on the comparative analysis of the light curves 
of BCC during the early rise to outburst maximum with those of the 
quiescent state of the binary. 
Our results show that a one-armed spiral structure develops in the 
accretion disk of EX Dra on the rise to maximum.
Section\,\ref{dados} describes the data analysis. 
The results are presented in section\,\ref{results} and 
discussed in section\,\ref{discuss}.

\section{Data Analysis} \label{dados}

Our data are $R$ band light curves obtained with the 0.9-m James Gregory
Telescope at the University of St. Andrews in 1995. 
The reader is referred to BCC for a detailed description of the data set
and of the reduction procedures. 
We collected three eclipse light curves of EX Dra on the rising branch
of the 1995 November outburst (eclipse cycles 7812, 7813 and 7817), 
two days after the onset of the outburst and two days before the
outburst maximum (outburst D of BCC; see their Fig.\,1).
The quiescent state is represented by four light curves (eclipse cycles 
8007, 8008, 8011 and 8013), obtained on 1995 December 27-28, about 10
days after the end of the subsequent outburst (outburst E of BCC).
In order to improve the signal-to-noise ratio and to reduce the
influence of flickering, the individual light curves were combined to
produce average eclipse light curves for the two observed states.  
For each light curve, we divided the data into phase bins of 0.003 cycle
and computed the median for each bin. The median of the absolute 
deviations with respect to the median for each bin was taken as the
corresponding uncertainty. The light curves were phase folded according
to the sinusoidal ephemeris of BCC,
\[
T_{mid} = {\rm HJD}\; 2\,448\,398.4530(\pm 1) + 0.209\,936\,98(\pm 4)\,E +
\]
\begin{equation}
+ (8.2 \pm 1.5) \times 10^{-4} \; \sin \left[ 2\pi \frac{(E-968)}{7045}
\right] \; d \;\;\; .
\label{efem}
\end{equation}

Out-of-eclipse brightness changes are not accounted for by the eclipse
mapping method, which assumes that all variations in the eclipse light 
curve are due to the changing occultation of the emitting region by the
secondary star.
Orbital variations were therefore removed from the light curves by fitting
a spline function to the phases outside eclipse, dividing the light curve
by the fitted spline, and scaling the result to the spline function
value at phase zero. This procedure removes orbital modulations 
with only minor effects on the eclipse shape itself.

We applied eclipse mapping techniques (Horne 1985; Rutten, van Paradijs
\& Tinbergen 1992b; Baptista \& Steiner 1993) to our light curves
to solve for a map of the disk brightness distribution and for the 
flux of an additional uneclipsed component in each case. 
The uneclipsed component accounts for all light that is not contained 
in the eclipse map in the orbital plane (i.e., light from the 
secondary star and/or a vertically extended disk wind). The reader is
referred to Rutten et~al. (1992b) and Baptista, Steiner \& Horne (1996) 
for a detailed description of and tests with the uneclipsed component. 
For the reconstructions we adopted the default of limited azimuthal 
smearing of Rutten et~al. (1992b), which is better suited for recovering
asymmetric structures than the original default of full azimuthal 
smearing (see Baptista et~al. 1996).
Simulations that show the ability of the eclipse mapping method to 
reconstruct asymmetric structures such as spiral arms are presented
and discussed by Baptista et~al. (2000a).
As our eclipse map we adopted a grid of $75 \times 75$ pixels
centered on the primary star with side 2~R$_{\rm L1}$, where
R$_{\rm L1}$ is the distance from the disk center to the inner
Lagrangian point. The eclipse geometry is defined by the mass ratio 
$q$ and the inclination $i$. We adopted the parameters of BCC, $q=0.72$ 
and $i=85\degr$, which correspond to an eclipse width of the disk center 
of $\Delta\phi= 0.1085$. These combination of parameters ensure that 
the white dwarf is at the center of the map.

The statistical uncertainties of the eclipse maps were estimated with a
Monte Carlo procedure (e.g., Rutten et al. 1992). 
A set of 20 artificial light curves is generated, in which the data 
points are independently and randomly varied according to a Gaussian
distribution with standard deviation equal to the uncertainty at that
point.  The light curves are fitted with the eclipse mapping algorithm to
produce a set of randomized eclipse maps. These are combined to produce
an average map and a map of the residuals with respect to the average,
which yields the statistical uncertainty at each pixel.

\section{Results} \label{results}

The data and the model light curves are shown in the left-hand panels of
Fig.\,\ref{fig1}. 
Horizontal dashed lines indicate the uneclipsed component in each case.
The middle panels show the corresponding eclipse maps in the same
logarithmic grayscale. 
The maps in the right-hand panels show the asymmetric part of the maps
in the middle panels and are obtained by calculating and subtracting
azimuthally-averaged intensities at each radius.
%
%

The uneclipsed fluxes are 1.45 and 1.6 mJy, respectively, for the 
light curves in quiescence and during the rise. 
The uneclipsed component in quiescence is mostly due to the contribution 
of the secondary star to the $R$ band flux, and is comparable to the
quiescent mid-eclipse flux.
The larger uneclipsed flux during the rise is possibly the result of
additional emission from a (variable) vertically-extended disk wind 
component, which becomes non-negligible during the outbursts 
(see Baptista \& Catal\'an 2000).

The light curve in quiescence shows a flat-bottomed asymmetric eclipse 
that is the result of the almost total occultation of a bright source at
disk center and of a faint compact bright spot at disk rim. The spot in 
the eclipse map lies along the gas stream trajectory and yields a 
quiescent disk radius of $R_q = 0.51\pm0.02 \;R_{L1}$, in good agreement 
with the value measured by BCC.
The width of the eclipse in the rise light curve is the same as in
quiescence but the eclipse is skewed towards later phases, indicating
the occultation of an asymmetric brightness distribution in the 
trailing side of the disk.
The resulting eclipse map has a symmetrical, centered brightness 
distribution, and an asymmetric structure in the form of a spiral arm
that is clearly visible in the right-hand panel of Fig.\,\ref{fig1}.
The eclipse maps obtained with the Monte Carlo procedure consistently
show a similar spiral structure.
The asymmetric arc contributes about 22 per cent of the total flux of
the eclipse map.

In order to trace the distribution in radius and azimuth of the 
asymmetric structures, we divided the eclipse maps in azimuthal slices
(i.e., `slices of pizza') and computed the distance at which the 
intensity is a maximum for each azimuth. We restricted the search to
radii $R > 0.2\;R_{L1}$ in order to avoid being affected by the central 
brightness source. The results are plotted in Fig.\,\ref{fig2} as a
function of binary phase (binary phases increase clockwise in the 
eclipse maps of Fig.\,\ref{fig1} and phase zero coincides with the 
inner Lagrangian point L1).
The lower panel shows the dependency of the maximum intensity with 
binary phase, while the upper panel gives the radial position of the
maximum intensity as a function of binary phase (only the regions for
which the radius of maximum intensity is larger than $0.2\;R_{L1}$ are
shown). The error bars were derived via Monte Carlo simulations with the
eclipse light curves (section \ref{dados}).
%
%

There are clear differences between the asymmetric structures in 
quiescence and on the rise to maximum.
The quiescent bright spot is a well defined, compact source with narrow
azimuthal ($\Delta\phi\simeq 30\degr$) and radial ($\Delta R= 0.04\; 
R_{L1}$) extents. On the other hand, the asymmetric structure on the rise 
map is extended both in azimuth ($\Delta\phi \sim 180\degr$) and radius
(from $R\simeq 0.2$ to $0.43\;R_{\rm L1}$), showing a spiral pattern 
in which the radius of the maximum intensity continuously increases
with binary phase.
An estimate of the disk radius at the rising stage can be obtained from
the outer radius of the spiral arm, $R_r= 0.43\; R_{L1}$. 

We computed equivalent Keplerian velocities along the spiral structure
by assuming $M_1= 0.75 \pm 0.15 \; M_\odot$ and $R_{\rm L1}= 0.85\;R_\odot$
(BCC). We obtain velocities in the range $600-800 \; {\rm km\, s^{-1}}$, 
larger than the velocities ($400-600\; {\rm km\, s^{-1}}$) inferred for the
spirals seen close to outburst maximum (Joergens et~al. 2000).
Since the eclipse map and the Doppler tomogram correspond to different
outburst stages, we are not able to distinguish whether the different
velocities are the consequence of different radii for the spirals, 
or if they are the signature of sub-Keplerian velocities in spiral 
shocks -- as seen in IP Peg (Baptista et~al. 2000a).

\section{Discussion} \label{discuss}

Figures \ref{fig1} and \ref{fig2} show that the spiral arm is brighter
and closer to disk center than the quiescent bright spot. 
The ratio of the intensity of the outer parts of the spiral arm 
and that of the quiescent bright spot ($\simeq 3.5$, lower panel of
Fig.\,\ref{fig2}) is much larger than the expected increase in 
accretion luminosity owing to the reduction in radius (for a constant
mass accretion rate), $L_{acr}(R_r)/L_{acr}(R_q) = R_q/R_r = 1.2$.
This discrepancy indicates that the mass accretion rate on the rise to
maximum is larger than the mass inflow at the quiescent bright spot by
a factor of $\simeq 3$\footnote{ 
In fact, this is a lower limit, since the bulk of the accretion
luminosity at the smaller $R_r$ radius should move to shorter wavelengths
due to the higher effective temperature, making the ratio of intensities
in the R-band $I_\nu (R_r)/I_\nu (R_q) < L_{acr}(R_r)/L_{acr}(R_q)$.}.
Moreover, if the outer radius of the spiral arm can be assigned to the
disk radius, the comparison of $R_q$ and $R_r$ indicates that the
accretion disk shrinks at the onset of the outburst. 
This result is in agreement with the predictions of the MTI model 
(Warner 1995), and suggests that the outburst in EX Dra is driven by 
a burst of enhanced mass-transfer from the secondary star.

We now turn our attention to the nature of the observed asymmetric 
structure.
It is hard to understand the observed one-armed spiral structure in 
terms of tidally-induced spiral shocks. First, because the accretion disk
radius at the corresponding outburst stage (again assumed to be of the
same order of the outer radius of the spiral arm, $R_r= 0.43\; R_{L1}$) 
is significantly smaller than the range in radius ($0.56-0.75 \;R_{L1}$)
required to excite spiral shocks strong enough to be observed (Steeghs 
\& Steele 1999). 
Secondly, because the tidal effect of the secondary star should, in
principle, lead to the formation of a {\em two}-armed spiral structure
(e.g., Yukawa, Boffin \& Matsuda 1997; Armitage \& Murray 1998), 
instead of the observed one-armed spiral.

An alternative explanation for the spiral structure in terms of 
enhanced gas stream emission might be advocated, although the position
of the spiral arm is clearly not consistent with the location of the gas
stream trajectory (see Fig.\ref{fig1}). Inconsistently low mass ratios
($q\simlt 0.1$) would be required to bring the gas stream trajectory 
in closer agreement with the observed spiral pattern. If valid, this
alternative would also indicate an MTI origin for the outbursts of EX Dra.

A third explanation may be devised if we note that outbursting 
accretion disks might depart from the geometrically thin disk 
approximation. 
At the high inclination of EX Dra, this may have important effects 
on the eclipse maps.
It may be possible that the observed spiral 
pattern is the combination of emission from a bright spot/stream
running along the edge of the disk and the (apparently) enhanced 
emission from the far (with respect to the secondary star) side
of a flared accretion disk.
This scenario makes the bright spot on the rise a factor of 3 brighter
than the quiescent bright spot, also in agreement with the predictions
of the MTI model.

The observed spiral structure bears resemblance with the transient 
arc-shaped asymmetric structures seen in 2-D and 3-D numerical 
simulations of accretion flows, at the onset of mass transfer (Makita,
Miyawaki \& Matsuda 1998; Makita \& Matsuda 1998) or at the start of
an instability-driven outburst (R\'o\.zyczka \& Spruit 1993).
In this regard, useful constraints on future numerical simulations
attempting to reproduce the observed spiral pattern may be obtained
by noting that the individual light curves combined to produce the 
average rise light curve cover a time interval of 5 binary cycles. 
Their eclipse shape is the same under the uncertainties. Hence,
the spiral structure seems to be stationary on a reference frame 
co-rotating with the binary and stable for a time length of, at least, 
5 binary orbits.

Finally, we note that, contrary to the usual assumption, inside-out 
outbursts are not necessarily inconsistent with the MTI model, 
provided that the burst of mass transfer leads to the formation of
a spiral structure well inside the primary lobe, that expands 
thereafter both inwards and outwards under the effects of mass and 
angular momentum redistribution to form a fully developed accretion 
disk.

We thank Emilios Harlaftis for helpful discussions and comments which
stimulated the writing of this paper.
This work was partially supported by the PRONEX/Brazil program through
the research grant FAURGS/FINEP 7697.1003.00. RB acknowledges financial 
support from CNPq/Brazil through grant no. 300\,354/96-7.
MSC acknowledges financial support from a PPARC
post-doctoral grant during part of this work.

\clearpage

\centerline{\psfig{figure=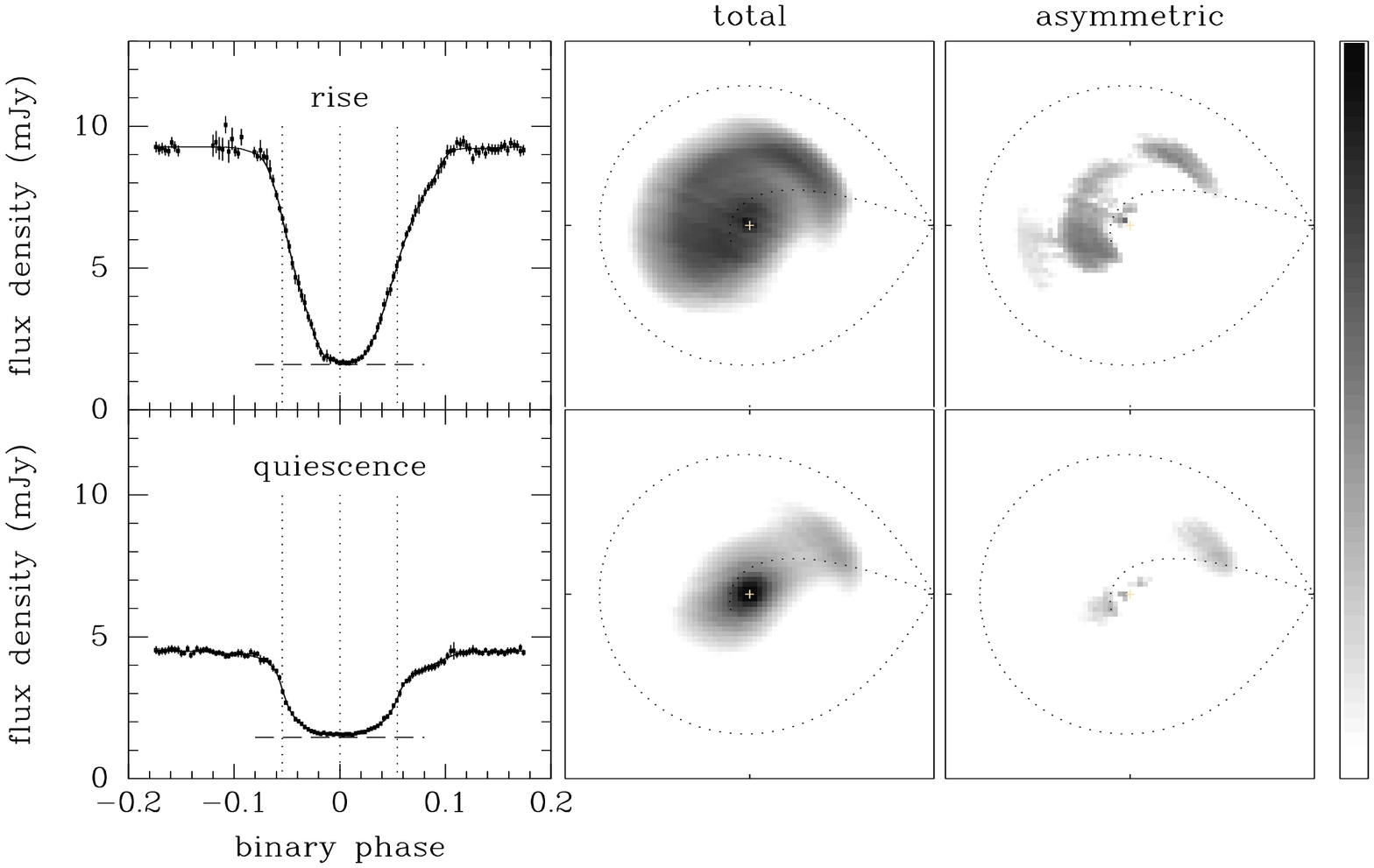,angle=-90,width=18.5cm,rheight=12cm}}
\figcaption[fig01.ps]
    { Left: Data (dots with error bars) and model (solid lines)
	light curves of EX Dra on the rise to maximum and in quiescence.
	Vertical dotted lines mark mid-eclipse and the ingress/egress times
	of the white dwarf. Horizontal dashed lines indicate the uneclipsed
	component in each case. Middle: eclipse maps in a logarithmic grayscale.
	Right: the eclipse maps of the middle panel after subtracting their
	symmetric part; these diagrams emphasize the asymmetric structures.
	Brighter regions are indicated in black; fainter regions in white.
	A cross mark the center of the disk; dotted lines show the Roche 
	lobe and the gas stream trajectory; the secondary is to the right 
	of each map and the stars rotate counterclockwise. \label{fig1}}

\centerline{\psfig{figure=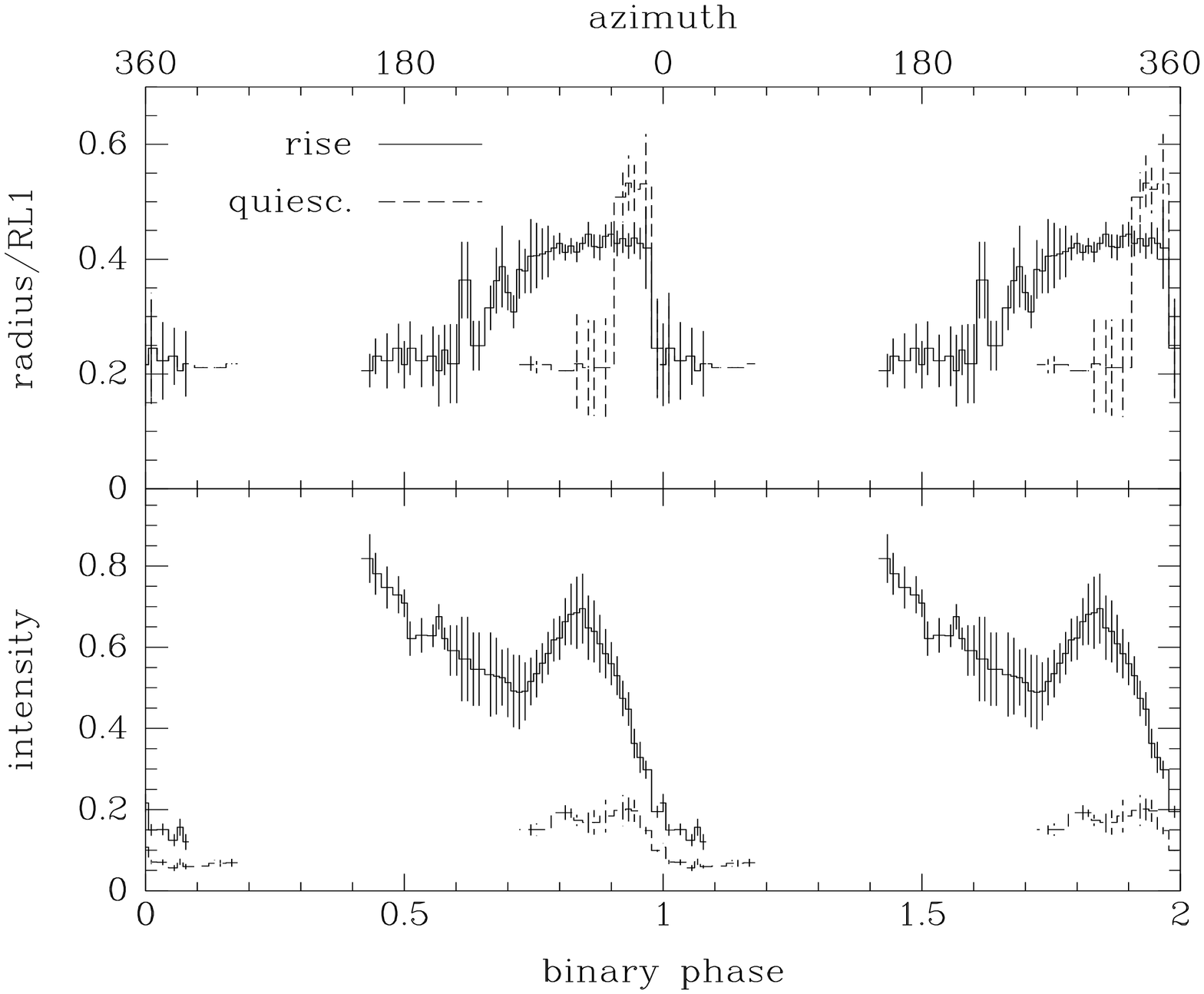,angle=-90,width=14cm,rheight=11cm}}
\figcaption[fig02.ps]
    { The dependency with binary phase of the maximum intensity and
	radius at maximum intensity, as derived from the eclipse maps on the
	rise (solid) and quiescence (dashed). The intensities are plotted in 
	an arbitrary scale. \label{fig2}}


\begin{thebibliography}{}

\bibitem {2} Armitage, P. J. \& Murray, J. R. 1998, \mnras, 297, L81
\bibitem {6} Baptista, R. \& Catal\'an, M. S. 2000, \mnras, submitted
\bibitem {8} Baptista, R., Harlaftis, E. T. \& Steeghs, D. 2000a, \mnras, 
		314, 727
\bibitem {4} Baptista, R., Catal\'an, M. S. \& Costa, L. 2000b, \mnras, 
		in press (astro-ph/0002382) [BCC]
\bibitem {10} Baptista, R. \& Steiner, J. E. 1993, \aap, 277, 331
\bibitem {12} Baptista, R., Steiner, J. E. \& Horne, K. 1996, \mnras, 282, 99
\bibitem {14} Hawley, J. F., Balbus, S. A. \& Winters, W. F. 1999, \apj,
		518, 394
\bibitem {16} Horne, K. 1985, \mnras, 213, 129
\bibitem {18} Hellier, C., et al. 2000, \mnras, in press (astro-ph/0001430)
\bibitem {20} Joergens, V., Spruit, H. C. \& Rutten, R. G. M. 2000, \aap, in
		press (astro-ph/0002302)
\bibitem {22} Makita, M., Miyawaki, K. \& Matsuda, T. 1998, preprint
		(astro-ph/9809003)
\bibitem {24} Makita, M. \& Matsuda, T. 1999, in proc. International Conference
		on Numerical Astrophysics, eds. S. M. Miyama et~al. (Boston: Kluwer
		Academics), 227 (astro-ph/9807237)
\bibitem {26} R\'o\.zyczka, M. \& Spruit, H. C. 1993, \apj, 417, 677
\bibitem {28} Rutten, R. G. M., Kuulkers, E., Vogt, N. \& van Paradijs, J.
		1992a, \aap, 254, 159
\bibitem {30} Rutten, R. G. M., van Paradijs, J. \& Tinbergen, J. 1992b, \aap,
		260, 213
\bibitem {32} Savonije, G. J., Papaloizou, J. \& Lin, C. 1994, \mnras, 268, 13
\bibitem {34} Sawada, K., Matsuda, T. \& Hachisu, I. 1986, \mnras, 219, 75
\bibitem {36} Steeghs, D., Harlaftis, E. T. \& Horne, K. 1997, \mnras, 290, L28
\bibitem {38} Steeghs, D. \& Stehle, R. 1999, \mnras, 307, 99
\bibitem {40} Stehle, R. 1999, \mnras, 304, 687
\bibitem {42} Vogt, N. 1983, \aap, 129, 29
\bibitem {44} Warner, B. 1995, Cataclysmic Variable Stars, Cambridge
		Astrophysics Series 28 (Cambridge: Cambridge University Press)
\bibitem {46} Yukawa, H., Boffin, H. M. J. \& Matsuda, T. 1997, \mnras, 
		292, 321

\end{thebibliography}
\end{document}